\documentclass[%
 aps,
 prl,
amsmath,amssymb,
reprint,
]{revtex4-1}

\usepackage{graphicx}
\usepackage{dcolumn}
\usepackage{bm}

\usepackage[utf8]{inputenc}
\usepackage{mathptmx}
\usepackage{multirow}
\usepackage{braket}

\usepackage[english]{babel}
\usepackage{xcolor}
\colorlet{RED}{red}
\colorlet{BLUE}{blue}
\usepackage{dcolumn}
\usepackage{bm}
\usepackage[version=4]{mhchem} 
\usepackage{acronym}
\usepackage{adjustbox}
\usepackage{float}
\usepackage{tikz}
\usepackage{microtype} 
\usepackage{algorithm}
\usepackage{xcolor}
\usepackage{algpseudocode}

\usetikzlibrary{calc,shapes.geometric,decorations.pathmorphing,patterns}

\definecolor{background-color}{gray}{0.98}
\usepackage[margin=2.3cm,bmargin=1cm,footnotesep=1cm]{geometry}

\begin{document}

\preprint{}

\title{\textbf{
Quantum Information Harvesting with the Parallel Quantum Flow Algorithm
} 
}%

\author{Nicholas P. Bauman}
 \email{Contact author:nicholas.bauman@pnnl.gov}
 \affiliation{Physical and Computational Sciences Division, 
  Pacific Northwest National Laboratory, Richland, Washington, 99354, USA}

\author{Ajay Panyala}
\affiliation{Physical and Computational Sciences Division, 
  Pacific Northwest National Laboratory, Richland, Washington, 99354, USA}

\author{Chenxu Liu}
\affiliation{Physical and Computational Sciences Division, 
  Pacific Northwest National Laboratory, Richland, Washington, 99354, USA}

\author{Muqing Zheng}
\affiliation{Physical and Computational Sciences Division, 
  Pacific Northwest National Laboratory, Richland, Washington, 99354, USA}

\author{Meng Wang}
\affiliation{Department of Electrical and Computer Engineering, the University of British Columbia, Vancouver, BC V6T 1Z4, Canada}

\author{Karol Kowalski}
\affiliation{Physical and Computational Sciences Division, 
  Pacific Northwest National Laboratory, Richland, Washington, 99354, USA}
\affiliation{Department of Physics, 
  University of Washington, Seattle, Washington, 98195, USA}

\date{\today}

\begin{abstract}
The Quantum Flow (QFlow) algorithm provides a resource-efficient framework for describing correlated many-body systems on hybrid quantum-classical architectures. By enabling parallel utilization of quantum and classical resources, QFlow offers a scalable pathway toward simulations of realistic systems. In this Letter, we report a high-performance computing (HPC) implementation of the QFlow formalism based on a singles-and-doubles model. We demonstrate its performance for target spaces comprising 82 and 114 orbitals, where the flow includes all 
6 active electrons in 6 active orbitals type active spaces. In the largest QFlow simulations, we optimize 1.17 million wave function parameters using the equivalent of 12 qubits. Despite the modest qubit requirements of the underlying active-space problems, the method recovers over $95\%$ of the total correlation energy obtained with the coupled cluster singles and doubles (CCSD) approach for systems dominated by dynamical correlation effects, which remain challenging for existing quantum algorithms. We further show that the QFlow formalism retains high accuracy in extended basis sets with diffuse functions, highlighting its potential for realistic large-scale quantum chemistry simulations.
\end{abstract}

\maketitle


\paragraph{Introduction.---} 
The accurate description of realistic correlated many-body systems in physics, chemistry, and materials science remains one of the central challenges of modern computational science. Owing to the exponential complexity of many-body wave function methods, predictive simulations of chemically and physically relevant systems often exceed the capabilities of even state-of-the-art high-performance computing (HPC) infrastructures. Quantum computing offers a fundamentally different computational paradigm that may overcome the memory and scaling limitations of classical approaches. Yet, the practical realization of this promise critically depends on the pace of the transition from noisy intermediate-scale quantum (NISQ) devices 
\cite{preskill2018quantum,bharti2022noisy,bauer2020quantum,chen2023complexity}
to fault-tolerant quantum computing (FTQC) \cite{shor1996fault,preskill1998fault,katabarwa2024early}, and on the development of many-body formalisms capable of exploiting the emerging generations of quantum hardware. A particularly important regime is expected to emerge with the advent of quantum platforms supporting $\sim 100$ logical qubits, projected for the 2028–2029 timeframe. In this regime, two qualitatively distinct scenarios may arise. In the first scenario, logical-qubit fidelities and non-Clifford gate budgets become sufficiently advanced to enable quantum phase estimation (QPE) simulations for realistic electronic Hamiltonians. In the second scenario, early FTQC devices remain too resource-constrained to support QPE calculations for dense many-body Hamiltonians, despite the availability of logical qubits. While the first scenario would unlock the full potential of quantum computing within a limited, yet non-trivial, qubit number, the second scenario necessitates the development of new hybrid many-body frameworks capable of delivering predictive results under the constraints of early FTQC architectures.

Over the last decade, hybrid quantum-classical methods have emerged as a promising route for extending the predictive reach of quantum simulations beyond the capabilities of either paradigm alone. Among these approaches, the Quantum Flow (QFlow) formalism \cite{kowalski2018properties,kowalski2023quantum,kowalski2023sub} provides an adaptive framework for parallel quantum-classical computing that naturally evolves with available quantum resources. QFlow controls quantum entanglement by decomposing a correlated many-body system into coupled low-dimensional active-space problems linked through rigorously defined effective (downfolded) Hamiltonians. By harvesting quantum information associated with this partitioning, QFlow enables distributed workflows in which effective Hamiltonians are constructed on classical platforms, while active-space optimization problems are solved using quantum resources. This framework provides a scalable pathway toward massively parallel simulations of strongly correlated systems using diverse quantum solvers. In this Letter, we demonstrate the performance of a highly scalable QFlow implementation for realistic systems and basis sets using quantum simulators, showing that large wave-function parameter spaces can be treated efficiently with modest quantum resources while recovering a surprisingly large fraction of the correlation energy.



\paragraph{Quantum Flow Method.---}
The QFlow formalism provides a resource-efficient framework for approximating the variational optimization problem  
\begin{equation}
E = \langle\Phi| e^{-\sigma} H e^{\sigma} |\Phi\rangle \;,
\label{var1}
\end{equation}
where the anti-Hermitian cluster operator  
operates in the target space. For realistic many-body systems, the corresponding target space is generally too large for direct quantum simulation.  
To enable simulations with limited quantum resources, we assume that the dominant correlation effects can be captured by a collection of reduced-dimensional active spaces $\{{\cal A}_i\}_{i=1}^M$ embedded in the target space. Equivalently, the cluster operator is approximated as a combination of all
non-repetitive excitations/de-excitations included in $\sigma_{\rm int}(i)$
anti-Hermitian cluster operators corresponding to active spaces ${\cal A}_i$ $(i=1,\ldots,M)$
\begin{equation}
\sigma \simeq
{\widetilde{\sum}}_{i=1}^{M}
\sigma_{\rm int}(i) \;.
\label{var2}
\end{equation}
For a given active space ${\cal A}_i$, the operator $\sigma$ is partitioned as  
\begin{equation}
\sigma \simeq
\sigma_{\rm int}(i)+\sigma_{\rm ext}(i) \;,
\label{var3}
\end{equation}
which yields the energy functional  
\begin{equation}
E=
\langle\Phi|
e^{-\sigma_{\rm int}(i)-\sigma_{\rm ext}(i)}
H
e^{\sigma_{\rm int}(i)+\sigma_{\rm ext}(i)}
|\Phi\rangle \; .
\label{var4}
\end{equation}
Partition (\ref{var3}) is commonly employed in active-space CC formulations \cite{pnl93}.

Following Ref.~\cite{kowalski2021dimensionality}, we employ an order-$N$ active-space-specific Trotter decomposition to define the approximate energy functional
\begin{equation}
E(i)=
\langle\Psi_{\rm int}(i,N)|
H^{\rm eff}(i,N)
|\Psi_{\rm int}(i,N)\rangle \;,
\label{var5}
\end{equation}
where  
\begin{equation}
H^{\rm eff}(i,N)=
(P+Q_{\rm int}(i))
\left[G_i^{(N)}\right]^{-1}
H
G_i^{(N)}
(P+Q_{\rm int}(i))
\label{var6}
\end{equation}
with  
\begin{equation}
G_i^{(N)}=
\left(
e^{\sigma_{\rm ext}(i)/N}
e^{\sigma_{\rm int}(i)/N}
\right)^{N-1}
e^{\sigma_{\rm ext}(i)/N},
\label{var7}
\end{equation}
and  
\begin{equation}
|\Psi_{\rm int}(i,N)\rangle =
e^{\sigma_{\rm int}(i)/N}|\Phi\rangle \; .
\label{var8}
\end{equation}

The QFlow algorithm is therefore defined by a set of coupled active-space problems [Eq.~(\ref{var5})] associated with active spaces ${\cal A}_i$ $(i=1,\ldots,M)$. The total variational manifold is spanned by all nonredundant amplitudes entering the active spaces included in the flow. Because of the noncommutativity of the $\sigma$ operators and the use of finite-order Trotter approximations, the energies $E(i)$ associated with different active spaces are generally not identical.  
In the original QFlow formulation, the active spaces were ordered according to physically motivated criteria. 
In this letter, we report the average energy across all active spaces as the representative QFlow energy, as the systems studied exhibit mainly dynamical correlations. For systems with strongly dominant excitation character or uneven active-space contributions, other heuristics may be more appropriate. 
A central advantage of the  QFlow formulation is that the quantum resource requirements are determined by the largest active space in the flow rather than by the full target space. In this Letter, we employ the simplest Trotter approximation corresponding to the $N=1$ case.

\paragraph{Numerical Implementation.---}
The QFlow workflow, implemented in the ExaChem quantum chemistry software package, comprises four major components: (1) selection of active spaces, (2) assigning amplitudes to active spaces, (3) parallel execution of the downfolding procedure and active-space solutions, and (4) updating of the global pool of amplitudes. 

Let $N_o$ and $N_v$ denote the total number of occupied and virtual spatial orbitals in the parent problem with index sets $\mathcal{O} = \{0, \ldots, N_o - 1\}$ and $\mathcal{V} = \{N_o, \ldots, N_o + N_v - 1\}$, respectively. The unitary coupled cluster with singles and doubles (UCCSD) ansatz \cite{hoffmann1988unitary,unitary1,unitary2,kutzelnigg1991error}, 
applied to an active space of $n_{o}$ occupied and $n_{v}$ virtual orbitals, recovers all single and double excitations within that space. With the overall goal of recovering all single and double excitations of the all orbital parent problem, the minimal active space required consists of two occupied and two virtual orbitals. The coverage set for the minimal size active space $\mathcal{C}_{\min}$ is the collection of all $\binom{N_o}{2}$$\binom{N_v}{2}$ four-orbital tuples ($o_1,o_2,v_1,v_2$) and the number of active space evaluations per cycle of QFlow is $|\mathcal{C}_{\min}|$. For larger active spaces, a single active space covers more elements, and therefore a smaller number of active spaces ($<|\mathcal{C}_{\min}|$) is needed to cover all single and double excitations.

\begin{figure}
\caption{Coverage-Driven Active-Space Sampling Algorithm}
\label{alg:sampling}
\begin{algorithmic}[1]
\Require Occupied orbital set $\mathcal{O} = \{0, \ldots, N_o - 1\}$,
         virtual orbital set $\mathcal{V} = \{N_o, \ldots, N_o + N_v - 1\}$,
         active space size $(n_{o}, n_{v})$,
         orbital energies $\{\varepsilon_p\}$,
         fixed random seed $s$
\Ensure  Ordered list $\mathcal{S}$ of active spaces covering all singles and doubles

\State Construct the minimal coverage set:
\[
\mathcal{C}_{\min} =
\bigl\{(o_1,o_2,v_1,v_2)\mid
o_1,o_2\in\mathcal{O},\;
v_1,v_2\in\mathcal{V},\;
o_1<o_2,\;
v_1<v_2
\bigr\}
\]

\State Initialize $\mathcal{R} \leftarrow \mathcal{C}_{\min}$,\quad
       $\mathcal{S} \leftarrow \emptyset$,\quad
       $\mathrm{RNG} \leftarrow \textsc{Seed}(s)$

\While{$\mathcal{R} \neq \emptyset$}
    \State Sample $\widetilde{\mathcal{O}} \subset \mathcal{O}$,
           $|\widetilde{\mathcal{O}}| = n_{o}$, uniformly at random without replacement
           using $\mathrm{RNG}$
    \State Sample $\widetilde{\mathcal{V}} \subset \mathcal{V}$,
           $|\widetilde{\mathcal{V}}| = n_{v}$, uniformly at random without replacement
           using $\mathrm{RNG}$
    \State Compute coverage contribution:
    \[
        \widetilde{\mathcal{C}} =
        \bigl\{(o_1, o_2, v_1, v_2) \mid
        o_1, o_2 \in \widetilde{\mathcal{O}},\;
        v_1, v_2 \in \widetilde{\mathcal{V}} \bigr\} \cap \mathcal{R}
    \]
    \If{$\widetilde{\mathcal{C}} = \emptyset$}
        \State \textbf{continue} \Comment{Candidate covers no new excitations; discard}
    \EndIf
    \State Compute orbital energy difference:
    \[
    \displaystyle\Delta\varepsilon =
        \sum_{v \in \widetilde{\mathcal{V}}} \varepsilon_v
        - \sum_{o \in \widetilde{\mathcal{O}}} \varepsilon_o
    \]
    \State Append $\bigl(\widetilde{\mathcal{O}},\, \widetilde{\mathcal{V}},\,
           \Delta\varepsilon\bigr)$ to $\mathcal{S}$
    \State $\mathcal{R} \leftarrow \mathcal{R} \setminus \widetilde{\mathcal{C}}$
\EndWhile

\State Sort $\mathcal{S}$ in ascending order of $\Delta\varepsilon$
\State Verify that $\mathcal{S}$ recovers $\mathcal{C}_{\min}$ exactly
\State \Return $\mathcal{S}$
\end{algorithmic}
\end{figure}

\begin{figure}
    \centering
    \includegraphics[width=\linewidth,trim={5cm 3.5cm 3cm 7cm}, clip]{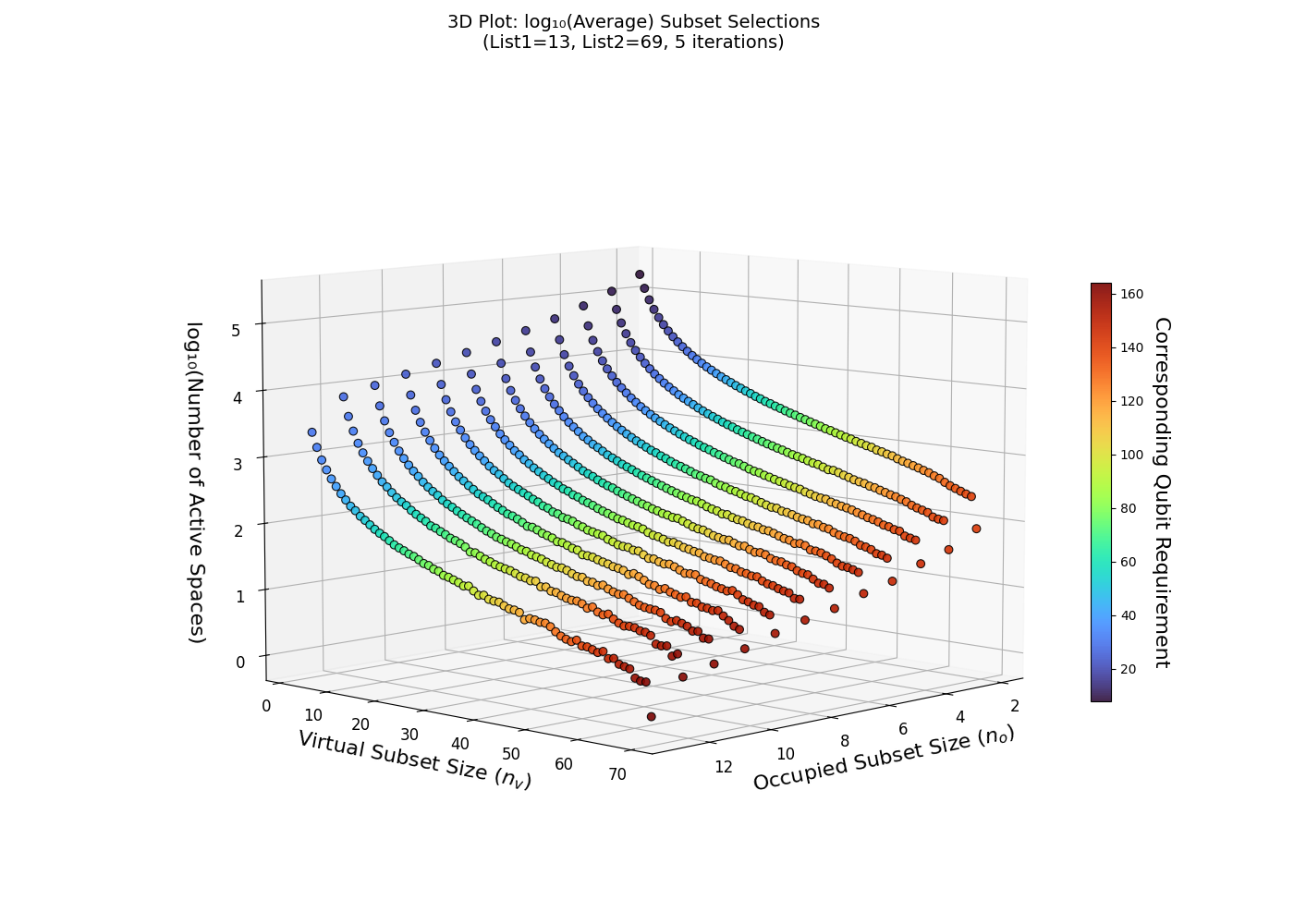}
    \caption{Number of active spaces for propane with the cc-pVDZ basis ($N_o$=13, $N_v$=69) determined for all combinations of active space sizes $(n_{o}, n_{v})$ using the active-space sampling algorithm. Active-space counts are color-coded to the corresponding qubit requirement [$2 \times (n_{o}+n_{v})$].}
    \label{fig:activespacecount}
\end{figure}

For active spaces with more than two occupied and two virtual orbitals, we implemented a coverage-driven active-space sampling algorithm to select the active spaces. As laid out in Figure ~\ref{alg:sampling}, we greedily construct a collection $\mathcal{S}$ of active spaces that cover all of $\mathcal{C}_{\min}$. Candidate active spaces are chosen by selecting $n_o$ orbitals from $\mathcal{O}$ to form $\widetilde{\mathcal{O}}$ and $n_v$ orbitals from $\mathcal{V}$ to form $\widetilde{\mathcal{V}}$. The coverage contribution from the candidate active space $\widetilde{\mathcal{C}}$ is checked against the remaining set $\mathcal{R}$ to see if there is at least one previously uncovered excitation. If so, then orbital energy differences are computed and added with the active space orbitals to the collection $\mathcal{S}$. Otherwise, it is discarded. This is repeated until all excitations are discovered. Figure \ref{fig:activespacecount} shows how the number of active spaces changes for the exemplary chemical system propane with the cc-pVDZ basis \cite{dunning1989gaussian}, which consists of 13 occupied and 69 virtual orbitals, using this sampling algorithm. The highest number of active spaces corresponds to the \{$n_o$,$n_v$\}=\{2,2\} active space, and when all orbitals are active, the single active space covers all single and double excitations.

A given amplitude can be associated with the solution of several different active spaces because they can overlap and share common orbitals. To ensure that an amplitude is updated only once per cycle, the update time is tied to a specific active space. After the active spaces are sorted in the sampling algorithm, excitations are assigned to the first active space (in energy order) that can reach them, ensuring each amplitude is updated exactly once in each active space solution.

After each active space is defined and the corresponding amplitude indices are assigned, the full Hamiltonian undergoes dimensionality reduction (downfolding) into that active space using the current global pool of amplitudes. The global pool is initially empty, but is progressively populated as active spaces are traversed and their solutions contribute updated amplitudes. Downfolding is performed via a DUCC (double unitary coupled cluster) calculation as implemented in ExaChem \cite{bylaska2024electronic}. The DUCC working equations are generated using the SymGen symbolic manipulator and are efficiently executed using the  Tensor Algebra for Many-body Methods (TAMM) C++ tensor library \cite{mutlu2023tamm}, which enables efficient CPU and GPU execution across heterogeneous hardware, from workstations to the world's largest supercomputers. The DUCC calculation yields an effective Hamiltonian, which is then passed to a quantum solver. In this implementation, we utilize NWQSim \cite{li2021svsim, li2020density, suh2024simulating, li2024tanq} to perform a VQE calculation \cite{peruzzo2014variational,mcclean2016theory,romero2018strategies,Kandala2017,kandala2019,izmaylov2019unitary,grimsley2019adaptive,mcardle2020quantum,Love2021} with a UCCSD ansatz. The optimized VQE parameters are returned to ExaChem, where they are matched to the active-space amplitude list to determine which parameters update the global pool. 

Each active-space problem is independent except for its coupling through the effective Hamiltonian, and therefore can be executed in parallel. Parallel execution is orchestrated using MPI process groups: process groups are assigned at initialization according to available resources, each group retiles/distributes data for efficient execution, performs downfolding and VQE calculation, updates amplitude in the global pool, then proceeds to the next active space in the list. Because the update order is effectively noncritical, as described below, there is no race bottleneck in the amplitude update. However, to ensure that each active space problem uses sufficient history in the global pool, we enforce a barrier at the end of each sweep across all active spaces before beginning the next cycle. The active space sweeps are repeated until convergence.

\paragraph{Results.---} We demonstrate the performance of the QFlow algorithm on two systems: water
(\ce{H2O}) and propane (\ce{C3H8}). For water, calculations were performed
using five basis sets ranging from cc-pVDZ, representing a 46-qubit problem, to
cc-pVQZ, representing a 228-qubit problem, with the 1\textit{s} core orbitals
of oxygen held uncorrelated in the QFlow calculation. For propane, all electrons
were correlated using the cc-pVDZ basis. The QFlow calculations for (\ce{H2O}) were carried out
using 280 cores to manage 140 process groups, each consisting of 2 cores, and 280 cores to manage 140 process groups, while the (\ce{C3H8}) calculation utilized 232 cores to manage 116 process groups. In all
calculations, the parent problem was represented by a series of active spaces
comprising 3 occupied and 3 virtual orbitals [$(n_o, n_v) = (3,3)$, corresponding
to 12 qubits].

\begin{figure}
    \centering
    \includegraphics[width=\linewidth, trim={1cm 0.5cm 0.5cm 0.5cm}, clip]{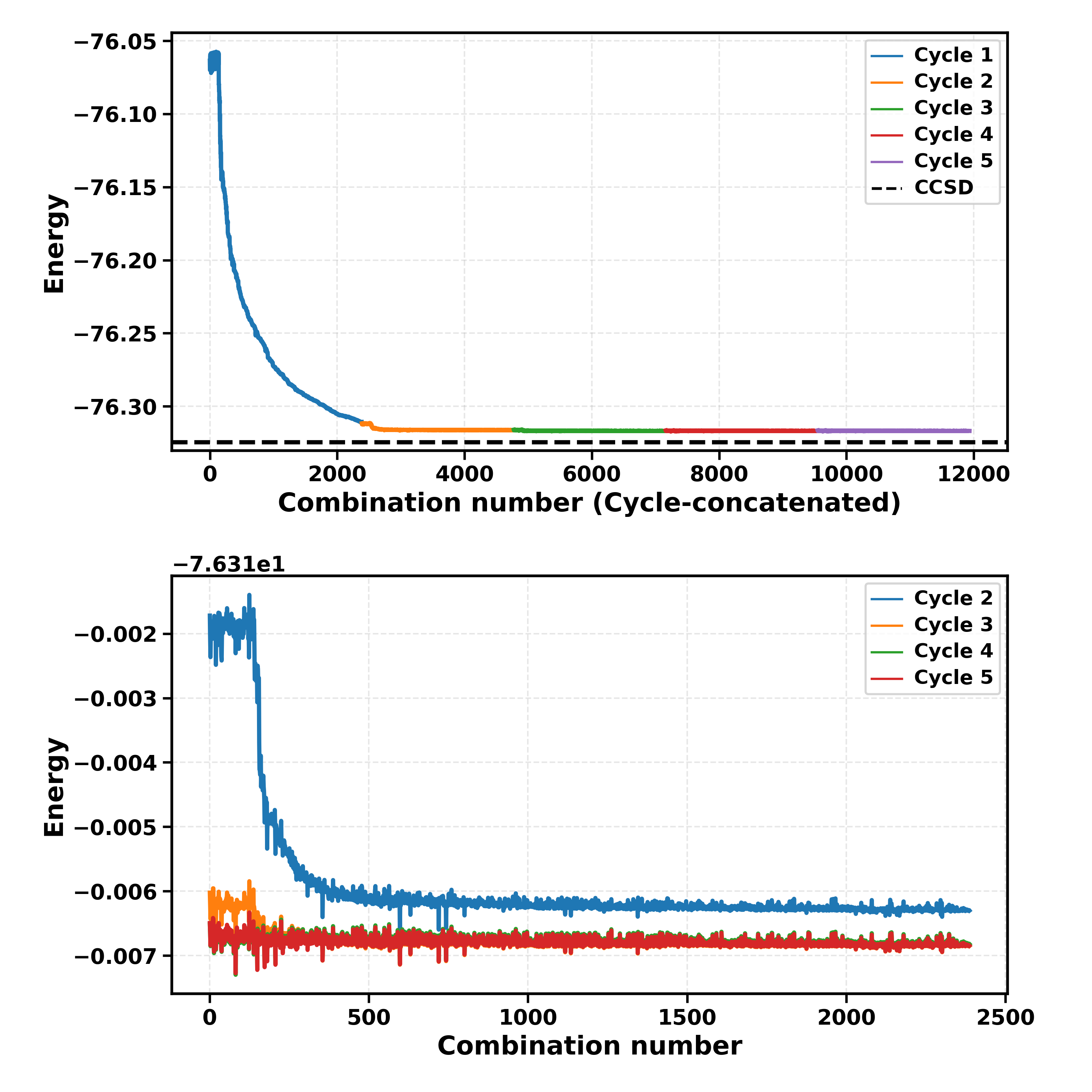}
    \caption{Top Panel: Energy profile of H$_2$O QFlow calculation using the cc-pVTZ basis. Bottom Panel: Energy profile of cycles 2-5 arranged by active spaces aligned in increasing orbital energy difference.}
    \label{fig:H2O}
\end{figure}

\begin{figure*}[ht!]
\includegraphics[width=\textwidth, trim={0.5cm 0.3cm 0cm 0cm}, clip]{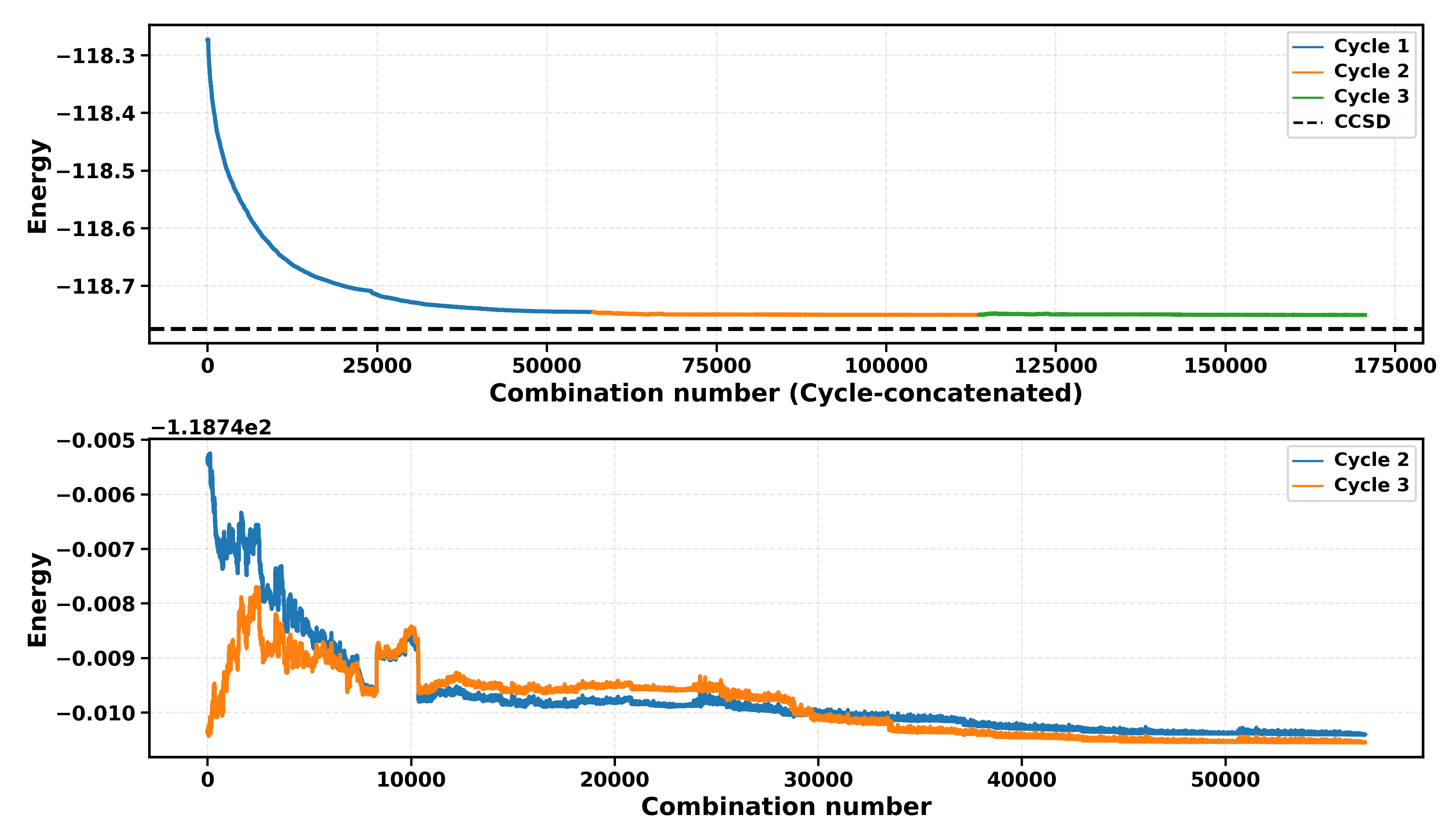}
\caption{\label{fig:propane} Top Panel: Energy profile of \ce{C3H8} QFlow calculation using the cc-pVDZ basis. Bottom Panel: Energy profile of cycles 2 and 3 arranged by active spaces aligned in increasing orbital energy difference.}
\end{figure*}

An exemplary QFlow energy profile is illustrated for \ce{H2O} with the cc-pVTZ
basis in Figure~\ref{fig:H2O}. The QFlow calculation replaces the 114-qubit
parent problem with cycles consisting of 2387 active spaces, each requiring only
12 qubits. The global pool of amplitudes is initially empty; consequently, as
the first 140 active-space combinations are dispatched in parallel, the
Hamiltonians correspond to the truncated bare Hamiltonian in the active space.
Each cycle, therefore, has an initialization period at the beginning of the cycle, but as the
active-space subproblems are solved, the global pool of amplitudes is updated, and
subsequent calculations incorporate information harvested from previously solved
active spaces. By the end of the first cycle, the QFlow calculation captures
95\% of the correlation energy relative to the all-orbital CCSD (CC with singles and doubles; Ref.\cite{purvis82_1910}) reference. The
amplitudes and energies converge rapidly in subsequent cycles: the energy change
in any single active space from the third to the fourth cycle is less than one
millihartree, and from the fourth to the fifth cycle, this is reduced to only
tens of microhartrees. This convergence behavior is consistent across all five
basis sets.

The QFlow algorithm enables quantum calculations that would otherwise require
prohibitively large qubit registers. As shown in Table~\ref{tab:basis_results},
calculation of \ce{H2O} with the cc-pVQZ basis would require 228 qubits,
whereas QFlow represents the problem using cycles of 10,319 twelve-qubit
subproblems. The ability of QFlow to capture correlations of large problems is
systematic: it consistently recovers 97\% of the correlation energy across all
five basis sets, with a slight increase as the basis set grows larger.

\begin{table}[ht]
    \centering
    \caption{QFlow results for \ce{H2O} with different basis sets,  }
    \label{tab:basis_results}
    \begin{tabular}{rcccc}
        \hline
        Basis & Qubits\footnote{Qubit requirement of the parent problem calculated as [$2 \times (N_{o}+N_{v})$].} & Cycle Size & QFlow (Mean) & \% Corr.\footnote{Percentage of correlation recovered by QFlow compared to canonical CCSD.} \\
        \hline
        cc-pVDZ       & 46   & 292 & $-76.2313$ & 96.8 \\
        aug-cc-pVDZ   & 80   & 1089 & $-76.2620$ & 97.1 \\
        cc-pVTZ       & 114  & 2387 & $-76.3168$ & 97.1 \\
        aug-cc-pVTZ   & 182  & 6412 & $-76.3259$ & 97.2 \\
        cc-pVQZ       & 228  & 10319 & $-76.3426$ & 97.2 \\
        \hline
    \end{tabular}
\end{table}

Simulating the full propane system requires 164 qubits for the cc-pVDZ basis, whereas the QFlow
algorithm replaces this resource cost with cycles of 56,860 active spaces to
optimize $1.17 \times 10^6$ unique parameters. The energy profile, shown in Figure~\ref{fig:propane}, is qualitatively
similar to that of \ce{H2O}. In the first cycle, the lowest-energy active space
recovers 94\% of the correlation energy. Subsequent cycles improve upon this
by an additional 1\%, reaching an average of 95\% by the third cycle, with each
active space obtaining submillihartree convergence compared to further iterations.

The ability of QFlow to consistently capture over 95\% of the correlation energy speaks to the robustness of the methodology. The remaining correlation can be systematically recovered through the hierarchy of approximations underlying the effective Hamiltonian. This includes incorporating higher-order many-body terms in the effective Hamiltonian and extending the commutator expansions at each downfolding step. Alternatively, the lower-order approximations can be made more accurate by employing larger active spaces. Larger active spaces reduce the extent of downfolding required while increasing the entanglement captured in each active space solve. Finally, to surpass singles and doubles accuracy, one must incorporate higher-order excitation parameters in both the downfolding and active space solvers. All of these aspects will be explored in future works. Our QFlow implementation is publicly available as part of the ExaChem repository at \url{https://github.com/ExaChem/exachem}~\cite{exachem_code}

\paragraph{Summary.---} 

We demonstrated that the QFlow formalism enables simulations of large target spaces in realistic basis sets while employing small 
$(n_o, n_v) = (3,3)$
active spaces. Despite the limited size of the underlying quantum problems, QFlow, combined with a UCCSD-VQE solver, recovers more than 95\% of the total CCSD correlation energy for systems dominated by dynamical correlation effects and described by target spaces approaching 100 orbitals. These results motivate several directions for further development especially in the context of strongly correlated states: selective identification of the minimal set of active spaces required for quantum-hardware simulations to uniquely characterize target states, machine-learning-guided optimization of active-space structures using VNet-based entanglement models \cite{liang2024effective,liang2025exploring}, and integration with advanced quantum solvers capable of capturing correlation effects beyond the UCCSD Ansatz. Since QFlow is rooted in exponential wave-function parametrizations, these extensions will also require efficient mappings of resulting quantum states onto unitary coupled-cluster representations to extract cluster amplitudes.
QFlow also provides an alternative pathway for utilizing large numbers of logical qubits through the execution of piecewise-disentangled shallow quantum circuits associated with independent active-space solvers. This framework offers a viable strategy for early fault-tolerant quantum computing regimes where error rates remain insufficient for deep-circuit implementations.

This material is based upon work supported by the ``Embedding QC into Many-body Frameworks for Strongly Correlated Molecular and Materials Systems''  project, which is funded by the U.S. Department of Energy, Office of Science, Office of Basic Energy Sciences, the Division of Chemical Sciences, Geosciences, and Biosciences (under FWP 72689) and by Quantum Science Center (QSC), a National Quantum Information Science Research Center of the U.S. Department of Energy (under FWP  76213). 
N.P.B and A.P. acknowledge  the support from 
Quantum Science Center (QSC), a National Quantum Information Science Research Center of the U.S. Department of Energy (under FWP  76213) and by the Quantum Algorithms and Architecture for Domain Science (QuAADS) Initiative, under the Laboratory Directed Research and Development (LDRD) Program at Pacific Northwest National Laboratory (PNNL). PNNL is operated by Battelle for the U.S. Department of Energy under Contract DE-AC05-76RL01830.


\begin{acknowledgments}

\end{acknowledgments}

\bibliography{references.bib}

\end{document}